\documentstyle[11pt,aaspp4]{article}

\def \m{\ifmmode M_\odot\else M$_\odot$\fi}

\def \gta {\mathrel{\vcenter 
     {\hbox{$>$}\nointerlineskip\hbox{$\sim$}}}} 
\def\av#1{\langle#1\rangle}

\def\deg{\ifmmode^\circ\else$^\circ$\fi} 

\begin{document}

\title{THE POLARIZED SPECTRUM OF APM 08279+5255}

\author{Dean C. Hines \& Gary D. Schmidt}
\affil{Steward Observatory, The University of Arizona, 933 
N. Cherry Ave., Tucson, AZ 85721} 
\affil{dhines@as.arizona.edu, gschmidt@as.arizona.edu} 
\authoraddr{Steward Observatory, The University of Arizona, 
Tucson, AZ 85721}

\author{and}

\author{Paul S. Smith} \affil{National Optical Astronomy 
Observatories,\altaffilmark{1} P.O. Box 26732} 
\altaffiltext{1}{Operated by the Association of Universities 
for Research in Astronomy, Inc., under contract with the 
National Science Foundation} \affil{950 N. Cherry Ave., 
Tucson, AZ 85726-6732} \affil{psmith@noao.edu} 
\authoraddr{National Optical Astronomy Observatories, P.O. 
Box 26732,} \authoraddr{950 N. Cherry Ave., Tucson, AZ 
85726-6732}

\begin{abstract}
We report the discovery of significant linear polarization 
($p > 1\%$) in the hyperluminous $z=3.87$ BALQSO 
APM~08279+5255.  The polarization spectrum is complex, with 
properties similar to those of other, lower redshift but 
more highly polarized BALQSOs.  The resonance emission lines 
are unpolarized while the absorption troughs show 
polarization similar to or higher than the continuum.  In 
particular, an apparent increase of polarization in the 
trough covering 1000$-$1030\AA\ (rest) supports the 
interpretation of this feature as a broad absorption 
component associated with O\,VI/Ly$\beta$ local to the QSO, 
as opposed to an intervening damped Ly$\alpha$ absorption 
system.

The elevated polarization in some of the absorption features 
implies that we view the scattered (polarized) spectrum 
through a sightline with less absorbing material than the 
direct spectrum.  Therefore, the complex structure of the 
polarization spectrum in this brilliant lensed BALQSO 
suggests that it will be an important laboratory for 
studying the structure of QSOs at high redshift.
\end{abstract}
 
 \vfill \eject

\section{INTRODUCTION}

Well into the third decade of investigation, the nature of 
QSOs is still uncertain.  However, expanded wavelength 
coverage and the advent of more powerful observing 
techniques have begun to place strong constraints on the 
environments and inner structure of Active Galactic Nuclei.  
In particular, optical spectro- and imaging polarimetry have 
revolutionized our understanding of Seyfert and radio 
galaxies, showing that orientation with respect to a dusty 
torus is a primary difference between broad- and narrow 
emission-line nuclei (e.g. Antonucci 1993; di Serego 
Alighieri, Cimatti, \& Fosbury 1994 and references therein).

The same techniques have revealed QSOs buried within dusty 
environments in several hyperluminous infrared galaxies 
(HIGs: e.g., Hines \& Wills 1993; Young et al.  1996).  These 
galaxies have QSO-like infrared luminosities and ``warm'' 
far-IR colors [$0.25 \le F_\nu(25\mu{\rm m})/F_\nu(60\mu{\rm 
m}) < 3$], indicative of dust heated by a strong UV 
continuum source (e.g. Low et al.  1989; Cutri et al.  
1994), but their QSO nuclei are hidden from direct view.  
Viewed from the direction of the scattering material which 
produces the polarization, the HIGs would be 
indistinguishable from typical luminous QSOs (Hines et al.  
1995; 1999; Hines 1999).

If HIGs represent QSOs viewed edge-on, and typical UV-excess 
QSOs are viewed more or less face-on, the identity of the 
``transition'' objects becomes an important issue.  A 
number of authors have suggested that these objects are 
represented by the UV-excess QSOs which show blueshifted 
resonant broad absorption lines (Weymann, Carswell, \& Smith 
1981).  The emission line properties of these Broad 
Absorption Line (BAL) QSOs, which number $\sim$10\% of all 
QSOs, are nearly identical to non-BALQSOs, and many 
investigators have concluded that all QSOs may contain BAL 
regions, but absorption is only evident along certain 
sightlines (e.g., Turnshek 1988; Weymann et al.  1991; Voit, 
Weymann, \& Korista 1993).  BALQSOs are more highly 
polarized, on average, than UV-excess QSOs, but less 
polarized than the HIGs (Schmidt \& Hines 1999, hereafter 
SH99).  Their polarized (scattered) spectra imply vantage 
points that just skim the attenuating region (Schmidt, 
Hines, \& Smith 1997; Ogle 1997; SH99).  Thus, the scattered 
and direct spectra probe different lines of sight to the 
nucleus.

Within the so-called Unified Scheme, QSOs, BALQSOs, and HIGs 
provide samples of objects viewed in three wedges of 
inclination, so intercomparison may potentially provide the 
basis for a detailed model of the central engine.  Except 
for the enormously luminous F10214+4724 (Rowan-Robinson et 
al.  1991; Eisenhardt et al.  1996), the sensitivity of {\it 
IRAS\/} limits the identification of HIGs to relatively low 
redshifts ($z \le 1$).  However, the space densities of 
luminous QSOs and BALQSOs peak at $z\sim 2$, making direct 
tests of this simple unified scheme more difficult.  
Fortunately, as for F10214+4724, gravitational lensing can 
amplify the flux, allowing access to higher redshift 
objects.

In this {\it Letter}, we examine the hyperluminous BALQSO 
APM~08279+5255.  Recently discovered in a search for red 
objects that might be distant carbon stars in the Galactic 
halo (Irwin et al.  1998), this $R = 15.2$, $z = 3.87$ 
object exhibits a BALQSO spectrum, warm far-IR colors, and a 
double image produced by lensing (Irwin et al.  1998; Lewis 
et al.  1998; Ledoux et al.  1998).  Even after correcting 
for a reasonable lensing amplification, this object ranks 
among the most luminous objects in the universe 
[$\log(L_{\rm IR}/L_\odot) \approx 14.5$], and provides a 
laboratory for testing the Unified Scheme, especially at 
high redshift.

\section{DISCOVERY OF POLARIZATION IN APM~08279+5255}

Linear polarization in APM~08279+5255 was initially detected 
on UT 1998 November 11 using the Two-Holer polarimeter 
operating in ``white'' light ($3200-8600$\AA) on the Steward 
Observatory (SO) 1.5~m telescope on Mt.  Lemmon (see e.g., 
Smith et al.  1992).  The value measured ($p\/ = 0.65\% \pm$ 
0.16\%, $\theta\/$ = 118\deg $\pm$ 7\deg) is similar to the 
polarization of a typical BALQSO (SH99), and sufficient to 
warrant detailed spectropolarimetric follow-up.  The 
additional observations were made on UT 1998 November 14 \& 
15 with the SO 2.3~m Bok telescope on Kitt Peak using the 
CCD Spectropolarimeter (Schmidt, Stockman, \& Smith 1992).  
Twelve $q$-$u$ sequences totaling 28,000~s were obtained 
during the two nights using a thinned, AR-coated Loral 
1200$\times$800 CCD. The instrument was equipped with a 
600~g~mm$^{-1}$ grating and a 3\arcsec\ wide EW slit to 
yield a resolution of $\sim$15\AA\ FWHM over the spectral 
range $4000-8000$\AA. Conditions were photometric throughout 
both nights, with seeing $1\farcs0-1\farcs5$ FWHM.

Observations of polarized standard stars allowed 
transformation of the polarization position angle to 
equatorial coordinates and the spectra of total flux were 
placed on an absolute scale through observations of spectral 
standard stars made each night.  The latter also allowed 
correction for discrete telluric absorption features.  Data 
from the two nights are entirely consistent, with no 
indication of variability, and so were combined.  Averaging 
over the entire spectrum yields a polarization of $p\/ = 
0.81\% \pm 0.02\%$ and $\theta\/$ = 110\fdg4 $\pm$ 0\fdg3; 
in good agreement with the broad-band measurement.

The Galactic latitude of the QSO ($b = +36\deg$) prompted a 
check for interstellar polarization (ISP) in white light 
using 3 stars within a radius of 10\arcmin.  These 
observations are summarized in Table~1.  The two fainter 
stars have similar polarizations and position angles, 
suggesting that they correctly represent the local ISP. The 
brighter star is probably not distant enough to probe the 
total ISP for this sightline and so was discarded.  The peak 
wavelength of the ISP was measured from spectropolarimetry 
of the star southwest of APM~08279+5255.  As indicated in 
the Table, the nearly 90\deg\ difference in $\theta\/$ 
between the QSO and nearby field stars causes the process of 
correcting for ISP to enhance the polarization of 
APM~08279+5255, with little change in position angle.  The 
polarization features observed in the spectrum of the object 
and discussed in the following sections further attest to 
the intrinsic polarization of the object.

\section{THE POLARIZED SPECTRUM OF APM~08279+5255}

The spectropolarimetry results for APM~08279+5255 are 
displayed in Figure 1.  Shown from top to bottom are: {\it 
a)\/} the polarization position angle $\theta$; {\it b)\/} 
the rotated Stokes parameter $q_\lambda^\prime$ for a 
reference frame aligned with the systemic position angle of 
102\deg; {\it c)\/} the rotated Stokes flux 
$q_\lambda^\prime \times F_\lambda$; and {\it d)\/} the 
total flux spectrum.  Strong emission lines from C\,IV, 
Ly$\alpha$, plus a weaker feature at 6900\AA\ associated 
with Si\,IV and OIV] are evident in the total flux 
spectrum\footnote{A typo in Irwin et al.  (1998) identifies 
the latter feature as Sulfur and O.}.  Also apparent are 
strong BALs associated with C\,IV, N\,V, and Ly$\alpha$, as 
are narrower absorption features previously identified as 
Mg\,II $\lambda$2800 from intervening material at lower 
redshifts.  We confirm a redshifted BAL feature associated 
with the C\,IV line, similar to that found in CSO~755 
(Glenn, Schmidt, \& Foltz 1994), and note that the feature 
also appears to be present in systems associated with N\,V, 
Ly$\alpha$, and O\,VI.

Considerable structure is seen in the percentage 
polarization spectrum.  In particular, the polarization {\it 
decreases} across the C\,IV, N\,V, and Ly$\alpha$ emission 
lines, but {\it increases} within some of the BAL systems.  
Despite its overall brightness, polarimetry of the BAL 
features in APM~08279+5255 is not easy.  The residual flux 
within the trough at 4900$-$5050\AA\ corresponds to 
$B\approx20.5$, and the degree of polarization averaged 
across this feature reaches only $p = 1.37\% \pm 0.15\% 
(\theta = 109.8\deg \pm 3.0\deg$).  That particular feature 
was interpreted by Irwin et al.  (1998) to be a damped 
Ly$\alpha$ system at $z = 3.07$, however the increased 
polarization apparent in our data implies that it arises 
within the QSO, not in an intervening system.  Ledoux et al.  
(1998) also point out that it is seen with equal equivalent 
width in the two lensed spectra.  In view of the 
emission-line redshift of the QSO, we identify the feature 
with broad absorption from O\,VI $\lambda\lambda$1032,1038 
and Ly$\beta$ (see e.g., Turnshek et al.  1996).  It is 
worth noting that the polarization within the observed 
Lyman continuum (4200$-$4500\AA) of the QSO is similar to 
that of the BALs ($p = 1.82\% \pm 0.24\%, \theta = 106.0\deg 
\pm 3.8\deg$) suggesting that the these two absorptions 
arises within the same clouds.

The polarized flux spectrum of APM~08279+5255 shown in 
Figure~1 is dominated by the continuum and BALs, testifying 
to a complete lack of polarization in the emission features.  
However, the equivalent widths of some of the BAL systems 
are reduced in the polarized spectrum as compared with the 
total spectrum, a fact which causes the increased degree of 
polarization at these wavelengths.  Coupled with a flat or 
slight increase in continuum polarization toward the blue, 
this behavior is typical of the polarized BALQSOs already 
studied by spectropolarimetry (Cohen et al.  1995; Schmidt, 
Hines, \& Smith 1997; Ogle 1997; Brotherton et al.  1997; 
SH99) and further substantiates the correlations summarized 
by SH99.  Previous studies have universally interpreted the 
polarimetry of BALQSOs in terms of multiple lines of sight 
to the nuclear source: a direct (unpolarized) view such as 
that which dominates the classic UV-excess (non-BALQSO) 
objects, and a polarized component produced by scattering 
akin to that which is prevalent in the narrow-line objects 
(e.g., Goodrich \& Miller 1995; Hines \& Wills 1995; SH99).  
The structured polarization spectrum of a BALQSO reflects 
the relative importance of these two components in, e.g., 
the emission and absorption features.  A second, scattered, 
sightline in APM~08279+5255 also naturally explains the 
residual flux observed in the BALs (see also Ledoux et al.  
1998).

For reasons of practicality, previous polarimetric studies 
of BALQSOs have emphasized the most strongly polarized 
examples: those with continuum polarizations $p \gta 2\%$ 
and trough polarizations typically above 5\%.  
APM~08279+5255, with values only half of these extremes, in 
one sense extends the correlations to the ``average'' BALQSO 
($\av{p}=1.26\%$; SH99).  The overwhelming similarity 
between the polarization characteristics of APM~08279+5255 
and the more strongly polarized BALQSOs indicates that they 
share a common basic nuclear geometry.  However, due to its 
high apparent brightness, APM~08279+5255 provides an 
exciting opportunity to further advance polarimetric probes 
into the inner structure of the nucleus.  The high redshift 
provides access to the Lyman continuum, and multiple BAL 
systems may enable comparison of different scattered 
sightlines.  The Mg\,II emission doublet, predicted to 
appear at a wavelength of 1.36$\mu$m, is accessible to 
modern IR spectrometers, so the BAL characteristics of the 
object for low-ionization species can be tested.  Finally, 
lensed systems like APM~08279+5255 (and H1413+117) may allow 
investigation of time delays and different amplifications of 
the various components.  Such observations may outstrip the 
capabilities of the 2~m telescope used for this work but are 
well within the grasp of instruments on 6$-$10 meter 
telescopes becoming available around the world.

\acknowledgments {We have benefited from illuminating 
discussions with C.B Foltz, G.F. Lewis, and F.J. Low.  We 
thank the anonymous referee for comments which improved the 
presentation.  Spectropolarimetry at Steward Observatory is 
supported by NSF grant 97-30792 to G.S. D.H. acknowledges 
additional support from NASA grants NAG 5-3359 and 
GO-05928.01-94A.}


\begin{deluxetable}{lccc}
\tablenum{1}
\tablewidth{0pt}
\tablecaption{POLARIMETRY OF THE APM~08279+5255 REGION}
\tablehead{\colhead{Object} & \colhead{$V$} & \colhead{$P$} & \colhead{$\theta$}}
\startdata
Star W\tablenotemark{a}		& 15.0	& $0.32 \pm 0.03\%$	
& $172\fdg4 \pm 3\fdg0$ \nl
Star SW\tablenotemark{b}	& 15.8	& $0.44 \pm 0.02\%$	
& $179\fdg8 \pm 1\fdg0$ \nl
Star E\tablenotemark{c}		& 11.7	& $0.10 \pm 0.02\%$	
& $145\fdg9 \pm 5\fdg0$ \nl
Adopted Interstellar (ISP)	&		
& $P_{\rm max}=0.42\%$\tablenotemark{d}	& 176$^\circ$ \nl
Measured APM~08279+5255 	& 		
& $0.81 \pm 0.02\%$	& 110$\fdg4 \pm 0\fdg3$ \nl
ISP-Corrected APM~08279+5255 	&		
& $1.12 \pm 0.02\%$	& $102\fdg4 \pm 0\fdg2$ \nl
\tablenotetext{a}{RA = 08:31:22.3; Dec = +52:44:59 (J2000)}
\tablenotetext{b}{RA = 08:31:38.5; Dec = +52:44:35 (J2000)}
\tablenotetext{c}{RA = 08:32:28.6; Dec = +52:45:34 (J2000)}
\tablenotetext{d}{$\lambda_{\rm max}\sim6500$\AA}
\enddata
\end{deluxetable}

\newpage

\begin{figure}[h]
\includegraphics{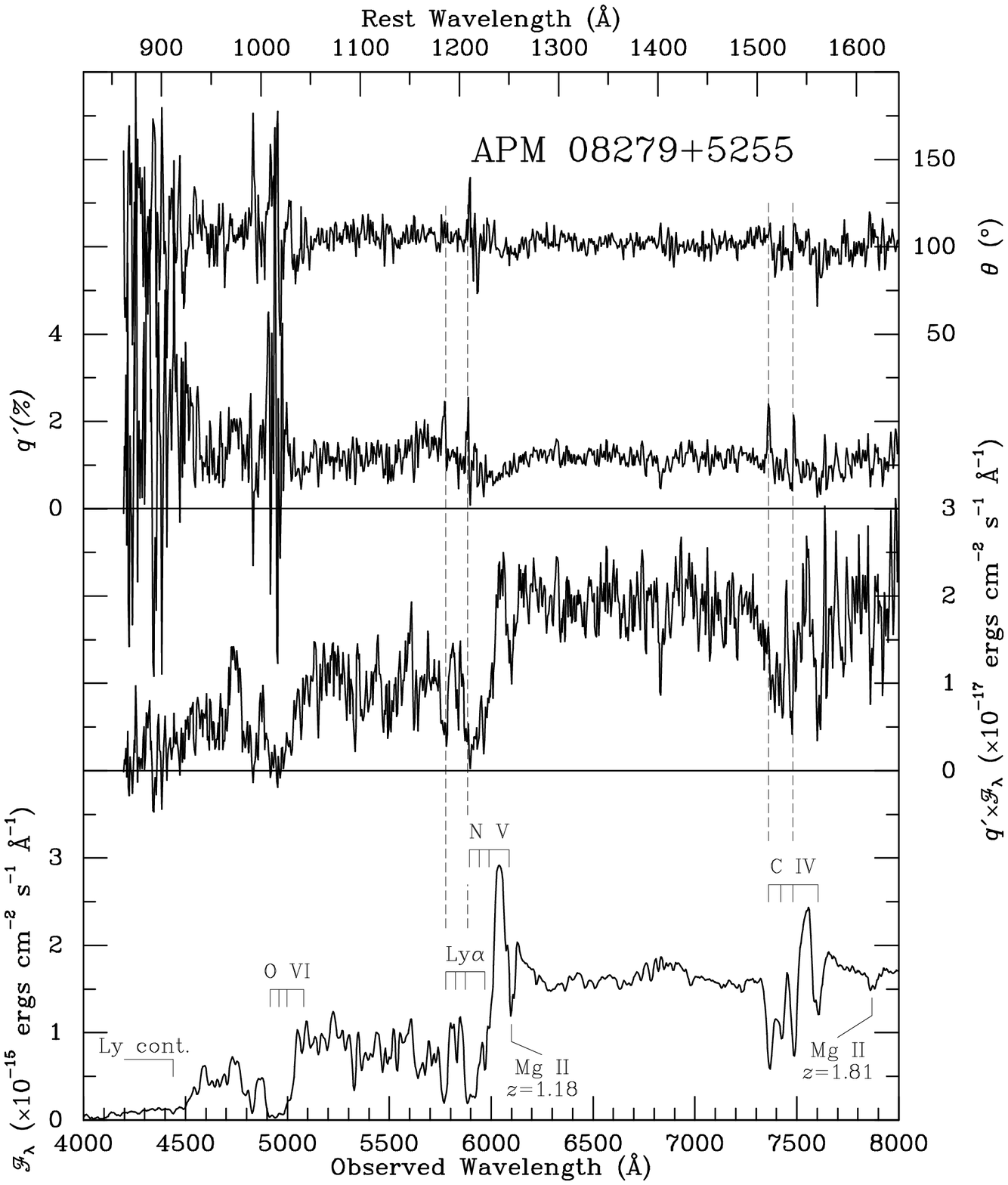}
\vspace{16.25truecm} 
\caption{
Spectropolarimetry of 
APM~08279+5255, corrected for Interstellar Polarization (see 
text): From top to bottom: {\it a)\/} the polarization 
position angle; {\it b)\/} the rotated Stokes parameter; 
{\it c)\/} the Stokes flux $q_\lambda^\prime \times$ 
F$_\lambda$; and {\it d)\/} the observed total flux density 
spectrum.  The $q^\prime$ parameter represents the total 
percentage polarization measured in a coordinate system 
aligned with the symmetry axis of the polarizing mechanism, 
but avoids the bias and peculiar error distribution 
associated with $p$.  Broad absorption features 
corresponding to those identified in C\,IV are marked for 
N\,V, Ly$\alpha$, and O\,VI/Ly$\beta$.  The Lyman continuum 
is also indicated.  Dashed vertical lines mark wavelengths 
at which the polarization in the C\,IV and Ly$\alpha$/N\,V 
absorption troughs increases significantly above the 
continuum.  These comprise two of the four principal BAL 
systems, indicated at +2000, $-$2830, $-$5250, and $-$7670 
km~s$^{-1}$ with respect to the emission-line redshift.
}
\end{figure}
 

\begin{references}
\overfullrule=0pt

\reference {} Antonucci, R.R.J. 1993, \araa, 31, 473

\reference {} Brotherton, M.S., Tran, H.D., Van Breugel, W., 
Dey, A. \& Antonucci, R.R.J. 1997, \apj, 487, L113
 
\reference {} Cohen, M.H., Ogle, P.M., Tran, H.D., 
Vermeulen, R.C., Miller, J.S., Goodrich, R.W., \& Martel, 
A.R. 1995, \apj, 448, L77

\reference {} Cutri, R.M., Huchra, J.P., Low, F.J., Brown, 
R.L., \& Vanden Bout, P.A. 1994, \apj, 424, L65

\reference {} di Serego Alighieri, S., Cimatti, A., \& 
Fosbury, R.A.E. 1994, \apj, 431, 123
	
\reference {} Eisenhardt, P.R., Armus, L., Hogg, D.W., 
Soifer, B.T., Neugebauer, G., \& Werner, M.W. 1996, \apj, 
461, 72

\reference {} Glenn, J., Schmidt, G.D., \& Foltz, C.B. 1994, 
\apj, 434, L47

\reference {} Goodrich, R.W., \& Miller, J.S. 1995, \apj, 
448, L73

\reference {} Hines, D.C. 1999, \apss, submitted

\reference {} Hines, D.C., Schmidt, G.D., Smith, P.S., 
Cutri, R.M., \& Low, F.J. 1995, \apj, 450, L1

\reference {} Hines, D.C., Schmidt, G.D., Wills, B.J., Smith, 
P.S. \& Sowinski, L.G., 1999, \apj, in press

\reference {} Hines, D.C., \& Wills, B.J. 1993, \apj, 415, 
82

\reference {} Hines, D.C., \& Wills, B.J. 1995, \apj, 448, 
L69

\reference {} Irwin, M.J., Ibata, R.A., Lewis, G.F., 
\& Totten, E.J. 1998, \apj, 505, 529 

\reference {} Ledoux, C., et al. 1998, \aap, 339, L77 

\reference {} Lewis, G.F., Chapman, S.C., Ibata, R.A., Irwin, M.J., 
\& Totten, E.J. 1998, \apjl, 505, L1 

\reference {} Low, F.J., Cutri, R.M., Kleinmann, S.G., \& 
Huchra, J.P. 1989, \apj, 340, L1

\reference {} Ogle, P.M. 1997, in Mass Ejection from AGN, 
ASP Conf.  Series, Vol.  128, N. Arav, I. Shlosman, \& R.J. 
Weymann, eds., p.  78
	
\reference {} Rowan-Robinson, M., et al. 1991, \nat, 351, 719 

\reference {} Schmidt, G.D. \& Hines, D.C. 1999, \apj, in 
press (SH99)

\reference {} Schmidt, G.D., Hines, D.C., \& Smith, P.S. 
1997, in Mass Ejection from AGN, ASP Conf. Series, Vol.  
128, N. Arav, I. Shlosman, \& R.J. Weymann, eds., p. 106

\reference {} Schmidt, G.D., Stockman, H.S., \& Smith, P.S. 
1992, \apj, 398, L57

\reference {} Smith, P.S., Hall, P.B., Allen, R.G., \& Sitko, M.L. 
1992, \apj, 400, 115

\reference {} Turnshek, D.A. 1988, in Proceedings of the QSO 
Absorption Line Meeting, C. Blades, D. Turnshek \& C. 
Norman, eds., 17

\reference {} Turnshek, D.A., Kopko, M.J., Monier, E., 
Noll, D., Espey, B.R., \& Weymann, R.J. 1996, \apj, 463, 110

\reference {} Voit, G.M., Weymann, R.J., \& Korista, K.T. 1993, 
\apj, 413, 95

\reference {} Weymann, R.J., Morris, S.L.,
Foltz, C.B., \& Hewett, P.C. 1991, \apj, 373, 23

\reference {} Weymann, R.J., Carswell, R.F., \& Smith, M.G. 1981, 
\araa, 19, 41

\reference {} Young, S., Hough, J.H., Efstathiou, A., Wills, 
B.J., Bailey, J.A., Ward, M.J., \& Axon, D.J. 1996, \mnras, 
281, 1206

\end{references}
\end{document}